# Direct magnetocaloric effect measurement technique in alternating magnetic fields

Aliev A.M.

*Amirkhanov Institute of Physics of Daghestan Scientific Center, RAS, Makhachkala 367003, Russia*

lowtemp@mail.ru

A method for direct measurement of the magnetocaloric effect (MCE) in alternating magnetic fields is offered. Main advantages of the method compared to a classical direct one are the high temperature sensitivity (better than $10^{-3}$ K); the ability of measuring of MCE in weak magnetic fields (by several tens oersteds and higher); the ability of measuring of MCE on small-sized samples ($1 \times 1 \times 0.01$ mm$^3$ and larger); the ability of measuring of MCE in alternating magnetic fields up to 50 Hz of frequency. The results on measurement of MCE on Gd and Ni-Mn-In Heusler alloy are reported.

## I. INTRODUCTION

The essence of the magnetocaloric effect (MCE) is the adiabatic temperature change (or magnetic isothermal entropy change) in a sample at the change of the external magnetic field due to the redistribution of internal energy between the system of spins and a crystal lattice. The MCE underlies an operation of a magnetic refrigerator of which design becomes a highly topical problem last twenty years [1-5]. Currently, dozens of prototypes of magnetic refrigerators are designed [6]. Along with well-studied gadolinium and some alloys [7], considered as advanced cooling materials at present, the efforts are under way to search new magnetocaloric materials [8-12].

A MCE value either is estimated by the direct measurement of a change in a sample temperature $\Delta T$ at application of the external magnetic field $\Delta H = H_f - H_s$ ($H_f$ is a finite value of the field, $H_s$ is the initial field) or is calculated indirectly using Maxwell relations from measurements of magnetization and/or heat capacity [13-14]. Weiss was the first to offer the direct measurement of MCE as earlier as 1921. These methods were considered in detail by Tishin and Spichkin in [14]. In materials with large MCE, the magnetic phase transition is accompanied generally by the structural transitions. Indirect methods in this case lead to faulty estimation of MCE as the Maxwell relations underlain the indirect methods are applicable for the second-order transitions. It often results to overestimated values [15]; therefore, the direct measurements appear to be more reliable. However, the direct measurement has a number of substantial disadvantages. First and foremost, it is a fairly large uncertainty (7-15 %) [14] and low temperature sensitivity. The last leads to necessity of large-sized samples what excludes out the possibility to study small-sized samples (ribbons, films, microwires). The MCE measurement in weak fields is also one of drawbacks of the classical direct measurement. As well as, the adiabaticity must be satisfied.

Over the last years, several different MCE measurement techniques, partially precluding the disadvantages of the conventional measurements, have been worked out [16-21]. The principle of most of these methods is the applying an alternating magnetic field to magnetocaloric material, which induce the temperature oscillations in the last. The alternating signal can be measured to higher accuracy than the DC one. As the offered methods are of high-sensitivity, a temperature change is failed to be measured directly by most of them and it is estimated in different manners from the temperature response of a sample to the magnetic field change.

## II. PRINCIPLES OF MCE MEASUREMENT

In the present work, we provide the detailed description of a technique for direct MCE measurement wherein the temperature change of a sample in the alternating magnetic field is measured directly. Using this technique, we have got the first results in 2009 [22]. The principle of the method is as follows. A studied sample is exposed to low-frequency alternating magnetic field

$$H = H_0 Sin\omega t,$$

where $H_0$ is the field amplitude, $\omega$ is the cyclic frequency. The field induces temperature oscillations in the sample

$$T = \pm T_0 |Sin(\omega t + \varphi)|,$$

where $\varphi$ is the phase shift of temperature oscillation in relation to the magnetic field change (Fig. 1). Signs ± denote a sample heating and cooling in relation to the mean temperature with increasing in field and correspond to direct and inverse magnetocaloric effects. The phase shift may come about by virtue of thermal resistance between the measured material and a thermocouple and also of a spin and spin-phonon relaxation rates in the sample. Temperature oscillations induce the variable thermo-electromotive force (*emf*) $E_{therm} = E_{therm0}|Sin(\omega t + \varphi)|$ in a differential thermocouple glued to the sample. The *emf* signal passes through a transformer preamplifier SR554 and is registered by a SR830 Lock-in Amplifier with high accuracy. The computer reads data from the lock-in and converts into the amplitude values of temperature oscillations.

Besides the temperature oscillations the alternating magnetic field induces the *emf* in the thermocouple wires loop, even when the MCE fails to be in the sample. A change in magnetic field direction results to changing of the sign of induced *emf* (blue dot line in the Figure). A sign of temperature oscillation is independent of the magnetic field direction. From a standpoint of MCE induction, the alternating magnetic field (black line in the Figure) can be shown as the variable magnetic field of one direction (cyan dash line in the Figure). The reference signal source generates the voltage at each increasing in the magnetic field. This implies that the frequencies of the reference signal, effective magnetic field, and temperature oscillations are equal to $2*\omega$, where $\omega$ is the angular frequency of an initial magnetic field (Fig. 1 depicts the periods of these signals). At the same time, the sign of induced *emf* changes with the changing of magnetic field direction. As it is evident from Figure, the frequency of this signal is 2 times lesser than that of reference signal, i.e. is equal to $\omega$. So, irrespective of *emf* occurred in thermocouple wire loop, this signal is not registered by the lock-in because the last measures a signal on the reference signal frequency only. Hence, the *emf* does not influence on MCE values. If the source of the magnetic field generates an oscillating magnetic field, but does not change the field direction, a temperature-dependent background signal is added to the measured signal. The temperature dependence of background is weak and caused by a small change of the input impedance with temperature, since the transformer preamplifier gain varies with the input impedance. A value of the background signal at given frequency increases with the magnetic field amplitude, because of increasing the rates of change in magnetic field and magnetic flux through the loop. The subtraction of the background signal from the measured one is necessary to obtain MCE true values.

**III. EXPERIMENTAL SETUP**

A measuring facility circuit diagram is drawn in Fig. 2. The alternating magnetic field is generated by means of an electromagnet and an externally controlled power supply. The control alternating voltage is applied to the power supply TDK Lambda Gen 100-7.5 from the generator output of the Lock-in Amplifier. To get the alternating magnetic field we connect the unipolar power supply output (TDK Lambda Gen) to the system which commutates the current in electromagnet coils every half-cycle. By this technique it is easy to obtain the weak alternating fields (up to 4 kOe). Using this source the MCE can be measured at 0.2-0.8 Hz frequency range of the field change. At large frequencies, the magnetic field amplitude strongly decreases because of the large inductance of electromagnet coils. Due to the same reason, there are problem to generate moderate and strong alternating magnetic fields by means of the electromagnet even at low frequencies of the field change. Adjustable Permanent Magnet Based Magnetic Field System 18 kOe serves as the source of moderate alternating magnetic field. This source generates the alternating magnetic field of 18 kOe amplitude and of frequency up to 0.8 Hz.

A sample mounting circuit is drawn in Fig. 3. The sample may be of any form but must have a plane surface whereon the junction of a differential chromel-constantan thermocouple is glued. The thermocouple is glued to a sample by Butvar-Phenolic Adhesive-2 providing a good thermal contact at temperatures up to 400 K, over that the high-temperature lacquer should be used. The differential thermocouple is prepared from wires of 25 μm in diameter. The thermocouple junction is done by an electric welding; the ends of wires are preliminarily flattened up to 3-4 μm in thickness in order to reduce the thermal inertia of the thermocouple. The mean temperature of the sample is measured by a copper-constantan thermocouple installed into a heat stock adjacent to studied sample. To eliminate an influence of the alternating magnetic field on mean temperature measurements a signal from the copper-constantan thermocouple is recorded for complete period of the field varying and then averaged. A heater is a constantan wire double wound around a copper-made cell. A change in sample temperature is provided by a thermal exchange gas $^4$He.

At such mounting of a sample, the heat capacity can be measured by the ac-calorimetry method. By selecting certain thickness and frequency of the periodic heating of the sample, the heat capacity can be defined using a formula

$$C = \frac{dQ}{2\pi T_{ac}}$$

where $dQ$ is the amount of heat periodically supplied to sample, $T_{ac}$ is temperature oscillations in the sample [23].

The precise determination of the gain factor of the lock-in + preamplifier system is one of the problem immediately appeared at measuring of the MCE. In a low-frequency range, this factor strongly depends both on the input impedance and on the frequency. For our SR830+SR554 system, the gain factor is defined by feeding the standard alternative signal at different input impedances within the range of which the resistance of the differential thermocouple changes.

The temperature sensitivity of the method is better than $10^{-3}$ K. This provides the possibility for measuring of MCE in the magnetic field of several tens of oersted in amplitude. The high temperature sensitivity and use of low thermal inertia thin thermocouples permits to measure the MCE on small-sized samples. A MCE measuring process in a wide temperature range from 100 K to 200 K takes a few hours. The requirement to design of special alternating magnetic sources is the only significant disadvantage of the method. The strong magnetic field sources, in particular, superconducting magnetic systems, don`t afford a fast change of the field.

For such magnetic sources, the method can be modified by loading and unloading thermal insert with the sample in the magnetic field with some constant frequency.

We use the described method for a few years and yield new attractive results in the research of rapidly quenched ribbons, microwires and bulk samples of Heusler alloys, and manganites of different compounds [24-31]. For testing the method, we use a well-studied material gadolinium Gd, which possesses one of the highest values of the magnetocaloric effect at room temperatures. The studied sample with 2x1x0.4 mm$^3$ sizes is cut from a single crystal.

## IV. MAGNETOCALORIC EFFECT RESULTS

Before starting the MCE-temperature measurements, it is necessary to ensure that the adiabatic condition is obeyed. The adiabaticity at conventional direct method is considered to be obeyed when a sample is in a high vacuum and a heat loss through thermocouple is minimized. The adiabaticity at this method is considered to be obeyed due to the high rate of the field change. Because of the low thermal inertia the thermocouple detects fast changes in temperature of the sample. As regards the sample temperature, it can be controlled by gluing a heater to the sample but then a mass and heat capacity of the heater must be taken into account at the calculation of the MCE absolute value. It is more easily to surround a sample by a thermal exchange gas. The satisfaction of the adiabatic conditions was tested measuring the field dependence of MCE in Gd with and without thermal exchange gas near the Curie point (Fig. 4, Inset). At the same frequency but large amplitude of the field change the rate of the field variation is higher. Therefore, a failure of the adiabaticity can be expected at small amplitudes and low frequencies of field change, when the sample can be time for the heat exchanging with surroundings. Results presented in Fig 4 exhibit the MCE values measured at the magnetic field change at 0.3 Hz and 0.8 Hz at high vacuum and with exchange gas. As it is clear from Figure, the data are precisely agree with each other. It indicates that the adiabatic conditions are satisfied well enough even in the case of an exchange gas.

The temperature dependences of MCE in Gd are shown in Fig. 4. A maximum MCE is observed at 291.7 K. A slight deflection on curves appears near 230 K (see Inset) due to a spin-reorientation transition [14]. As it is evident, the proposed method allows measuring the small values of MCE (thousandth of degree) with high relative accuracy. The results agree with literature data for Gd. But it should be noted that a wide spread of data on the direct measurement of the MCE in Gd (0.20-0.40 K/kOe) prevails in the literature [14].

As stated before, the phase shift between the changes of magnetic field and temperature of the sample can be because of the thermal resistance of a sample and the thermocouple, and the relaxation times of spin-phonon and spin system in the sample. Fig. 5 exhibits the temperature dependences of phase shift in Gd at different amplitudes of magnetic field changing. As it is seen, the phase shift has the pronounced features in the ferromagnetic-paramagnetic and the spin – reorientation transitions regions. With magnetic field increasing, the temperature dependence of phase shift is begun to reduce, while in the vicinity of phase transitions it almost does not depend on the field value. On the one hand, in the vicinity of magnetic phase transitions the magnetic subsystem is most sensitive to the external field, so decreasing of the phase shift between the change in the magnetic field and temperature change to be expected. On the other hand, because of fluctuations, relaxation processes strongly slow down near the Curie point. Therefore it is difficult to define the nature of the behavior of phase shift basing just on these results. Obviously, the reasons for such behavior cannot be explained by the thermal contact between the thermocouple and sample, since neither thermocouple nor glue and nor sample reveal such features of the thermal conductivity in the studied temperature range. The understanding of the nature of phase shift requires additional studies.

One of the problems appearing at measuring the MCE by the proposed approach is the detection of a temperature change sign. The lock-in measures the amplitude value of an alternating signal which inherently is positive. In fact, there is a method of determining the temperature change sign. The use of the SR830 Lock-in Amplifier or analogous detectors permits the detection of the temperature variation sign if both direct and inverse MCE happen in a sample. The SR830 Lock-in Amplifier includes two synchronous detectors whereon the reference signals are shifted by 90° relative to each other [32].

The Phase Sensitive Detector (PSD) output is proportional to $V_{sig} \cdot \cos(\theta)$, where $\theta = \theta_{sig} - \theta_{ref}$. $\theta$ is the phase difference between the signal and the lock-in reference oscillator. By adjusting $\theta_{ref}$ we can make equal $\theta$ to zero, in which case we can measure $V_{sig}$ ($\cos(\theta) = 1$). Conversely, if $\theta$ is 90°, there will be no output at all. A lock-in with a single PSD is called a single-phase lock-in and its output is to $V_{sig} \cdot \cos(\theta)$.

This phase dependency can be eliminated by adding a second PSD. If the second PSD multiplies the signal with the reference oscillator shifted by 90°, i.e. $V_{sig} \cdot \sin(w_L t + \theta_{ref} + 90°)$, its low pass filtered output will be

$$V_{psd1} = \frac{1}{2} \cdot V_{sig} \cdot V_L \cdot \sin(\theta_{sig} - \theta_{ref}),$$

$$V_{psd1} = \frac{1}{2} \cdot V_{sig} \cdot \sin\theta$$

Now we have two outputs, one proportional to $\cos\theta$ and the other proportional to $\sin\theta$. The first and second outputs can be defined as $X$ and $Y$, in way

$$X = V_{sig} \cdot \cos\theta \qquad Y = V_{sig} \cdot \sin\theta.$$

These two quantities represent the signal as a vector relative to the lock-in reference oscillator. *X* is called the 'in-phase' component and *Y* the 'quadrature' component. By computing the magnitude (*R*) of the signal vector, the phase dependency is removed:

$$R = \sqrt{X^2 + Y^2} = V_{sig}$$

*R* measures the signal amplitude and does not depend upon the phase between the signal and lock-in reference. A dual-phase lock-in, such as the SR830, has two PSD's, with reference oscillators 90° apart, and can measure *X*, *Y* and *R* directly.

Fig. 6 demonstrates the results of the MCE measurements in $Ni_{49.3}Mn_{40.4}In_{10.3}$ Heusler alloy, where both direct and inverse MCE take place in ferromagnetic to paramagnetic transition and magnetostructural transition regions, correspondingly. Fig. 6a presents measured *emf* of thermocouple $U_{thermocouple}$ and phase shift of *emf* relative to magnetic field. X- and Y-components of the thermocouple signal are shown in Fig. 6b. High-temperature direct MCE with a maximum at 318 K in the Figure is due to ferromagnetic-paramagnetic transition while a low temperature inverse MCE with maxima at 237 K and 226 K at heating and cooling respectively is due to a magnetostructural transition from high temperature austenite to low temperature martensite phase. When the direct - inverse MCE crossover occurs, the phase shift changes by more than 90°, as the sample temperature changes in opposite phase relative to the magnetic field change. The in-phase and quadrature components are positive in vicinity of the ferromagnetic - paramagnetic transition, i.e. the vector of the amplitude signal appears to be in the first quadrant. With temperature decreasing the amplitude signal vector proves to be in the third quadrant and respectively, the in-phase and quadrature components are negative vicinity of magnetostructural transition. The real values of MCE calculated from the measured $U_{thermocouple}$ and taking into account the sign change of MCE that is visible on phase shift and X-and Y-components curves, are shown in Fig.6c. The MCE value in moderate field ΔH=18 kOe are shown in Fig.6d.

The applicability of the Maxwell relations to magnetostructural first order phase transition, the comparison of direct and indirect MCE measurement results, including those in relation to the Heusler alloy, widely discussed in recent years [33-36]. Actually the classical direct measurement of MCE can give significant errors near the first order transition areas. Firstly, this is due to the width of temperature region, whereon the MCE is occurred, often doesn`t exceed 10 K, what technically precludes the accurate measurements. Secondly, when measuring the MCE, an entropy change caused by the structural transition may contribute into the measured value. The measurements of MCE can be done taking into account the structural contribution and excluding it. Therefore, there is a great divergence of results on MCE measurements at magnetostructural phase transitions. The offered approach measures only a magnetic contribution, as just a signal from this contribution changes in-phase with the change in the external magnetic field. The structural transition contribution into the MCE can reveal itself as one-time temperature change of the sample, but this signal is not recorded by the lock-in. In any case, direct measurements of the MCE in alternating magnetic fields are preferred over other methods because magnetocaloric properties are estimated in conditions similar to the conditions in which the magnetocaloric material will be in the real magnetic refrigerators.

**V. CONCLUSION**

We develop a new technique and the fully-automatic experimental set-up for the direct measurement of the magnetocaloric effect in alternating magnetic fields. The method possesses considerable advantages in comparison with the classical direct one:
- high temperature sensitivity;
- the ability of measuring of MCE in weak magnetic fields;
- the ability of measuring of MCE on small-sized samples;
- requires short time for measurements in a wide temperature ranges.

Up to now the researchers mainly focused on the study of MCE in moderate and high magnetic fields, because the main purpose of the studies is finding new materials for magnetic refrigerators. At the same time, MCE studies in weak fields, which are rare nowadays, can provide fundamental information about the magnetic properties of the materials. Therefore, we must extensively use of the possibility of the technique for studying the magnetic properties of materials in weak fields.

One of the main ability of the method, namely, the possibility for MCE studies in alternating magnetic fields up to 50 Hz, i.e. under conditions similar to those in real magnetic refrigerators. The dynamics methods, one of which is the proposed in the paper can be in demand when searching the materials for solid-state magnetic refrigerators. The first results on the study of the frequency dependence of MCE are obtained and will be published soon.

**ACKNOWLEDGMENTS**

Author thanks A.B. Batdalov and V.V. Koledov for valuable discussions and useful advices and I.K. Kamilov for support at all stages of the development of technique. This work was partly supported by the Branch of Physical Sciences of the Russian Academy of Sciences within the framework of the program "Strongly Correlated Electrons in Solids and Structures" and the RFBR (Grant Nos. 14-02-01177 and 12-02-96506).

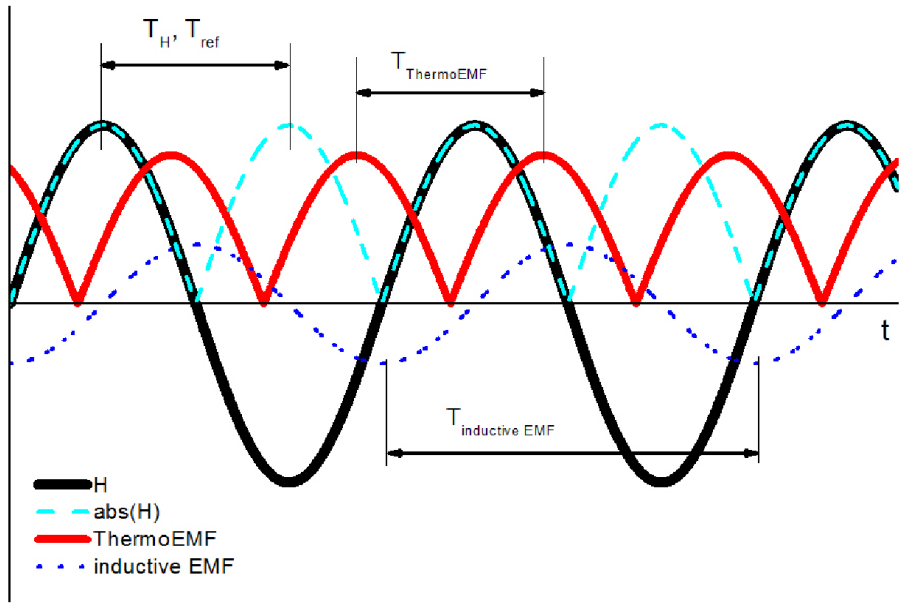

Fig.1. Alternating magnetic field (solid black line), effective magnetic field (cyan dash line), ThermoEMF induced by temperature oscillations in sample due to MCE (solid red line) and EMF induced by changing of the magnetic flux in the thermocouple wires loop (blue dot line). Periods of all signals are indicated.

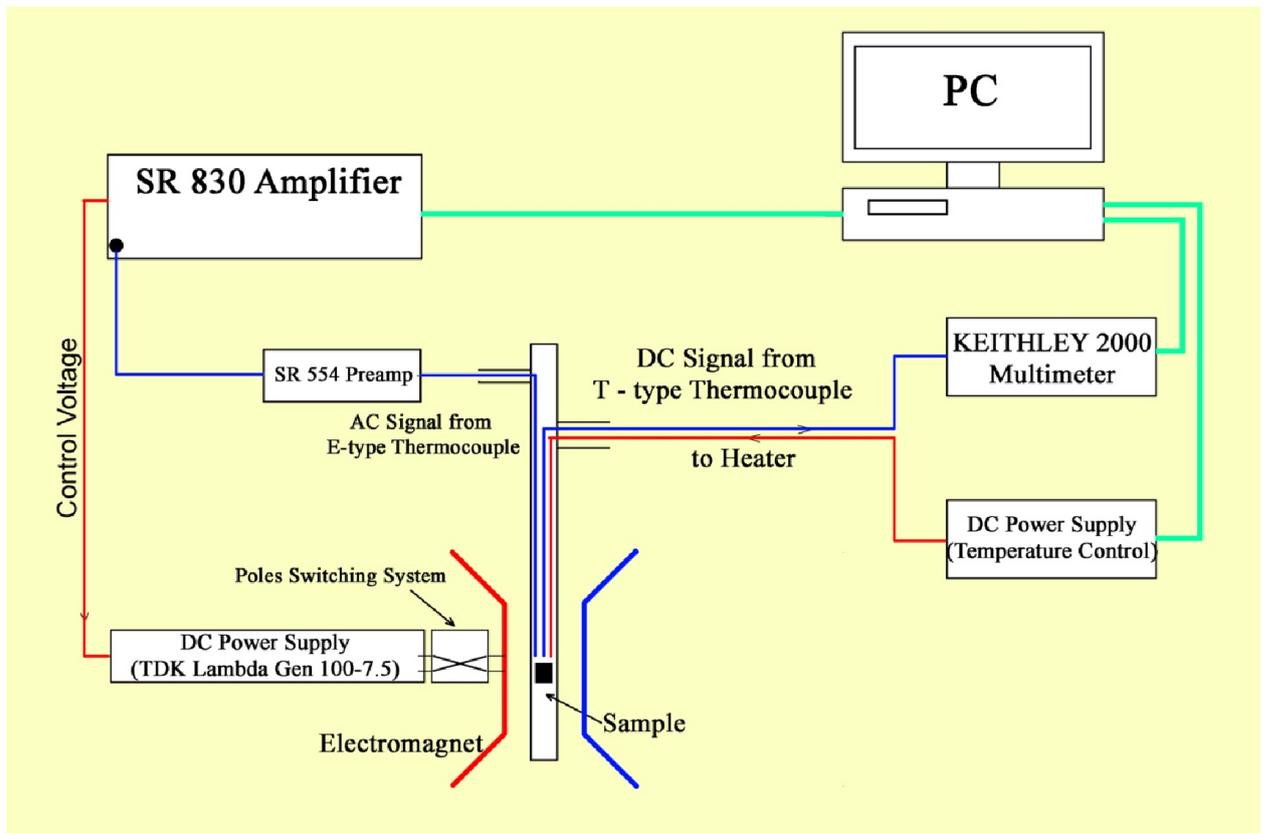

Fig.2. Circuit diagram of MCE measuring setup.

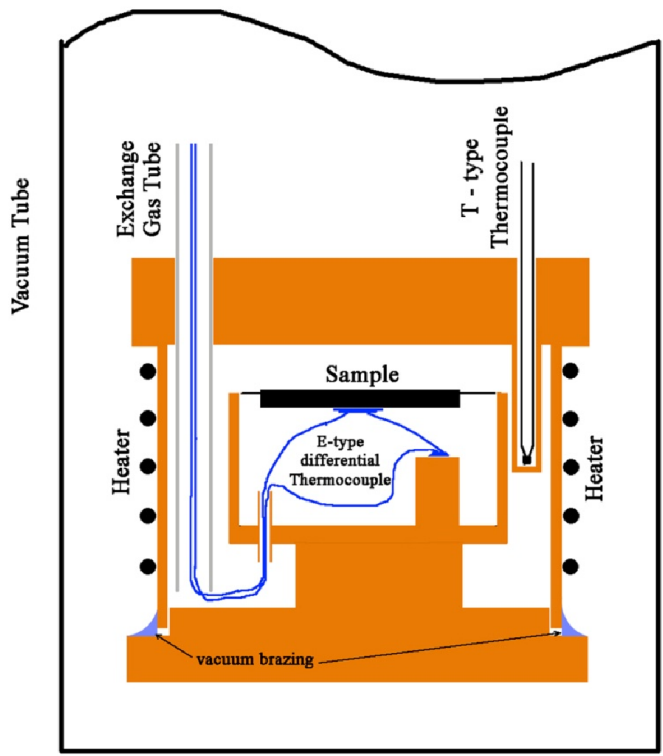

Fig. 3. Scheme of sample mounting.

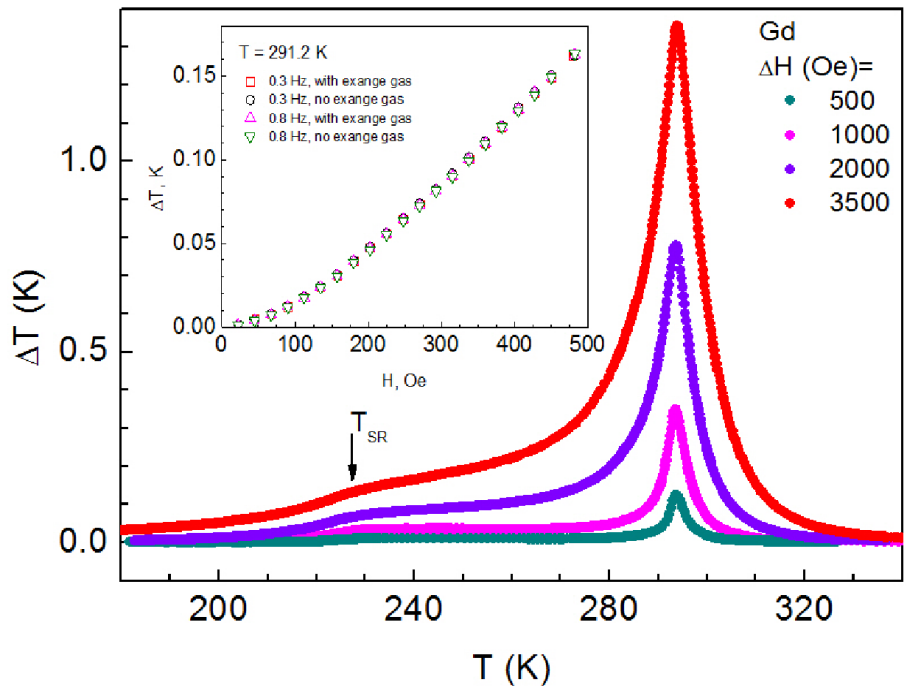

Fig. 4. MCE in Gd single crystal in weak magnetic fields. Inset – field dependence of MCE in and without exchange gas.

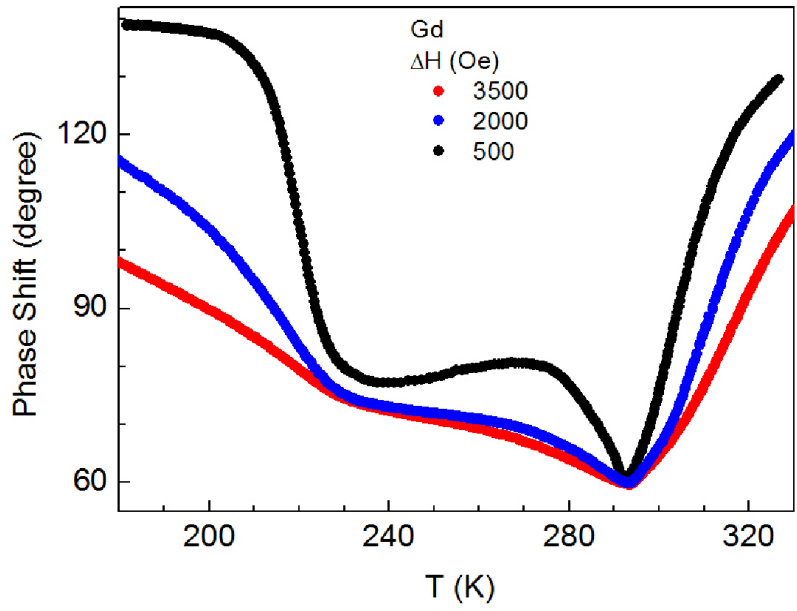

Fig. 5. Phase shift between the changes of magnetic field and temperature of the sample.

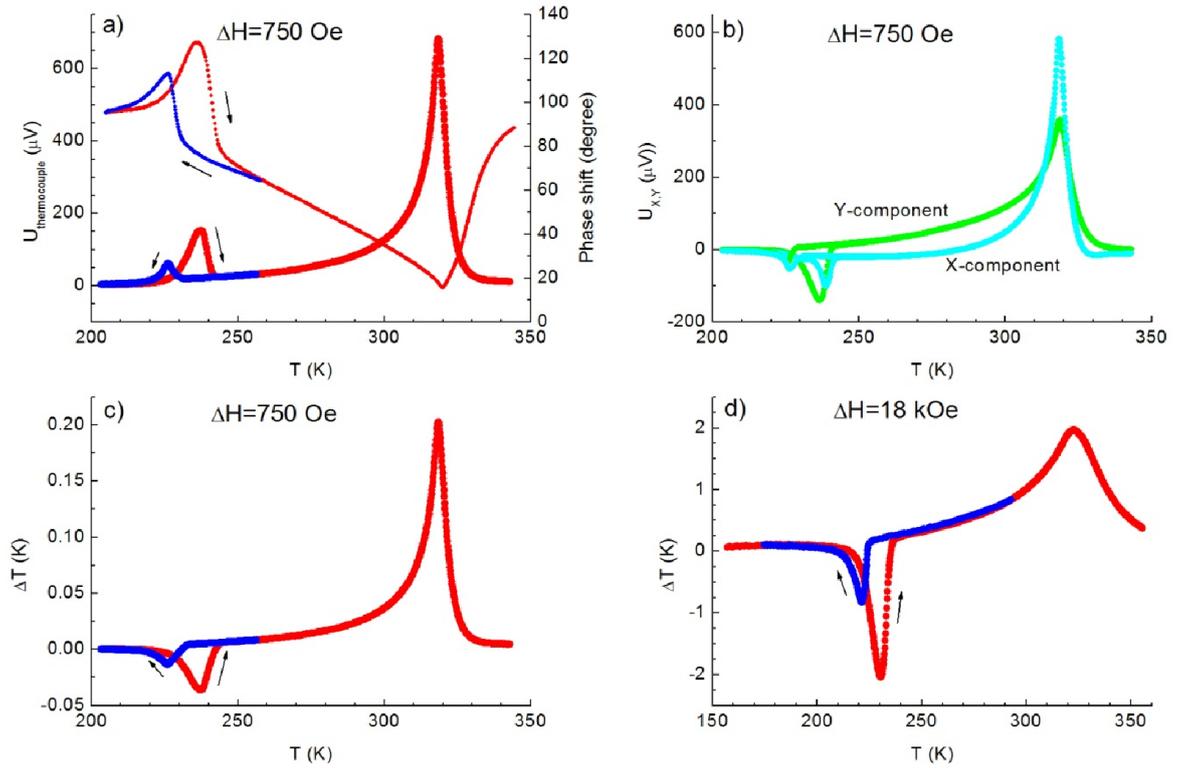

Fig. 6. MCE in $Ni_{49.3}Mn_{40.4}In_{10.3}$ Heusler alloy. a) gained amplitude signal from thermocouple and phase shift between magnetic field and temperature of sample; b) X- and Y- components of the measured signal; c) $\Delta T$ in $Ni_{49.3}Mn_{40.4}In_{10.3}$; d) $\Delta T$ in $Ni_{49.3}Mn_{40.4}In_{10.3}$ at $\Delta H=18$ kOe. The measurements were carried out in the heating and cooling regimes.